\documentclass[a4paper]{article}
\usepackage[affil-it]{authblk}
\usepackage{amssymb}
\usepackage{enumerate}
\usepackage{color}
\usepackage{amsmath}
\usepackage{graphicx}
\usepackage{longtable}

\usepackage{url}

\begin{document}

\title{The reach of commercially motivated junk news on Facebook}

%
%

\author{Peter Burger\textsuperscript{1} , Soeradj Kanhai\textsuperscript{2}, Alexander Pleijter\textsuperscript{1}, Suzan Verberne\textsuperscript{2}\thanks{\texttt{s.verberne@liacs.leidenuniv.nl}; Corresponding author}}
\affil{1. Department of Journalism and New Media, Leiden University Centre of Linguistics (LUCL), Leiden University, Leiden, the Netherland\\
2. Leiden Institute of Advanced Computer Science (LIACS), Leiden University, Leiden, the Netherlands}

\maketitle

\begin{abstract}
Commercially motivated junk news -- i.e. money-driven, highly shareable clickbait with low journalistic production standards -- constitutes a vast and largely unexplored news media ecosystem. Using publicly available Facebook data, we compared the reach of junk news on Facebook pages in the Netherlands to the reach of Dutch mainstream news on Facebook. During the period 2013-2017 the total number of user interactions with junk news significantly exceeded that with mainstream news. Over 5 Million of the 10 Million Dutch Facebook users have interacted with a junk news post at least once. Junk news Facebook pages also had a significantly stronger increase in the number of user interactions over time than mainstream news. Since the beginning of 2016 the average number of user interactions per junk news post has consistently exceeded the average number of user interactions per mainstream news post.\end{abstract}

\section{Introduction}
Social media and Facebook in particular have become a major gateway to news. Large numbers of people access news through social media, as shown by survey data from the Reuters Digital News Report~\cite{newman}. In the US, 45\% of respondents used social media for news consumption on a weekly basis, with Facebook being the leading source. Unfortunately, not all news spread by social media consists of high-quality, well-edited content. An important factor in the quality of Facebook’s news feed is the widely discussed presence of ‘fake news’ and clickbait on the platform.  ~\cite{mosseri,DeVito2017} and – possibly – due to the widely discussed presence of `fake news' and clickbait on the platform. Because, unfortunately, not all news spread by social media consists of high-quality, well-edited content. Over the last years there has been an alleged rise in low-quality, and even completely fabricated news on social media~\cite{Allcott2017}. This paper attempts to assess the kind of news that reaches a nation -- in this case: the Dutch -- of Facebook users.

The rise of so-called `fake news' has gained much attention in academia, government and the media in the last few years. In this paper we refrain from using the term `fake news' as an analytical concept, since it is imprecise and heavily politicized, encompassing connotations such as deceitful, false, and slanted~\cite{Nielsen2017,Tandoc2018,wardle,venturini}. Instead, we use the term junk news, focussing on a combination of content characteristics, production values, and types of producers. We study the money-driven, low-quality, highly shareable kind of content that is typically distributed on social media as clickbait. This genre frequently includes -- but is not limited to -- disinformation, i.e. completely fabricated or severely distorted information presented in news formats.

Since the 2016 US elections, interest in the `information disorder'~\cite{wardle} has boomed. Academic research has focussed on the nature of the problem, its impact on the audience, and ways to counter it (e.g.~\cite{Allcott2017,Roozenbeek2018,Vargo2018}). Journalists have identified individuals and organisations spreading disinformation for ideological or commercial reasons~\cite{silverman}. Government bodies, think tanks, and social network platforms (Facebook, Twitter) have produced reports about covert foreign influence operations (e.g.~\cite{powers,house}). Within this `junk news universe', research focuses predominantly on political content.

However, quantitative studies about the reach of junk news and disinformation are scarce. An extensive review of the literature on disinformation and social media, published in 2018~\cite{Tucker2018} highlights the prevalence of various kinds of disinformation as a research gap~\cite[p. 56]{Tucker2018}. In addition, it notes an over-emphasis on Twitter and a lack of studies using Facebook data~\cite[p. 60]{Tucker2018}, and mentions restrictions imposed by the social media platforms~\cite[p. 70]{Tucker2018}. 

The present study investigates the reach of junk news on Facebook. More specifically, we take the Netherlands as a case. The Dutch Facebook network is extensive: there are approximately 10.5 Million Facebook users in the Netherlands~\cite{Statista2018}, on a total population of 17 Million. We compared the reach and development of commercially motivated Dutch junk news on Facebook to the reach and development of Dutch mainstream news on Facebook. For the purpose of this study we define mainstream news as well-edited content, published by established news media. We collected 117 thousand Facebook posts published by 63 junk news pages and 20 mainstream news pages over a five-year period. With these data, we study the reach of junk news and mainstream news by measuring publication activity and user engagement. Publication activity is defined by the number of posts published by a Facebook page. User engagement is defined by the number of user interactions with the published posts.

Given the alleged rise of junk news and in light of Facebook measures to improve news feed quality, the objective of this study was (1) to assess the total reach of junk news on Facebook, compared to mainstream news, in terms of user engagement; and (2) to investigate how junk news develops over time, in terms of publication activity of the junk news producers' Facebook pages, and of user engagement with the published posts.

\section{Background}
\subsection{Junk news defined}
Scholars and journalists use various terms when they discuss news that is in some respects deceitful and/or unreliable. Our study concerns junk news. The bulk of the production of the junk news pages that we include consists of low-quality, sensational content. They frequently publish fabricated – i.e. completely fake – news, but as Venturini~\cite{venturini} argues, diffusion, not falsity is the point here: `[...] spread, rather than fakeness, is the birthmark of these contents that should be called ``viral news'' or possibly ``junk news'' for, just as junk food, they are consumed because they are addictive, not because they are appreciated.'~\cite[p. 3]{venturini}  

`Junk news' is also employed as a term by Oxford University's project on Computational Propaganda~\cite{howard,Howard2017,Narayanan2018} covering a wide range of news sources. These are rated as `junk news' if they tick at least three of the following five boxes: lack of professionalism (low to non-existent journalistic standards); sensationalist style (in-your-face visuals and headlines, strong emotional appeal); low credibility (low-quality sources, no fact-checking, false information, conspiracy theories); bias (hyper-partisan reporting); and forgery (outlets imitate both news formats and specific news brands, to pass off their fakes as genuine)~\cite[p. 2--3]{Narayanan2018}.

We adopt the term `junk news' from Narayanan et al.~\cite{Narayanan2018}, but adapt the definition in order to make it applicable to Dutch commercial junk news on Facebook. Studying social media use during elections, the Computational Propaganda (ComProp) papers focus on politically themed and motivated social media messages. The characteristics of these messages are partly similar to the Dutch source material we analyzed. They differ in their emphasis on ideology and falsehood. First, the outlets we tracked are not ideology-driven, but purely commercial. In addition, though falsehoods are frequent, they only occur in a minority of items; moreover, they mostly appear to be the consequence of low production standards rather than instances of intentional deception. Finally, none of the web sites in our sample deceptively imitates a respectable news brand, so the category `forgery' does not apply.

This leads us to the following characteristics that constitute our working definition of commercial junk news (henceforward: junk news):
\begin{itemize}
\item low journalistic quality (pre-packaged content, no added research and fact-checking);
\item produced by non-mainstream producers;
\item business model based on websites with advertising and Facebook pages pushing the sites' posts; 
\item goal is viral success;
\item frequently contains fabricated or heavily distorted messages;
\item frequent use of clickbait headlines.
\end{itemize}
The commercial incentive of the pages implies that pages that are ideologically motivated are not covered by our definition.

\subsection{Related work}

Most studies in this field have been conducted from a political disinformation perspective (e.g.~\cite{Allcott2017,Howard2018,humprecht2018fake}). Few address the reach of money-driven junk news. Moreover, most of the available data on commercial junk news have been published by investigative journalists. In a seminal exposé, Buzzfeed editor Silverman~\cite{silverman} showed that during the 2016 US elections, the most popular fake stories about politics outperformed the most popular mainstream news stories. 

Le Monde's fact-check and data journalism team Les D\'{e}codeurs analyzed 101 false claims, spread by 1001 web pages and videos. These claims were not all about politics: notably, health-related stories were among the most popular. Links to these pages and videos generated 4.3 million Facebook shares and some 16 million interactions (i.e. shares, comments, and likes). Three quarters of the false stories elicited more than 10,000 interactions each~\cite{senecat2017}. 
Using a sample of fact-checked news items, Vosoughi et al.~\cite{Vosoughi2018} found that news items labeled by fact-checkers as untrue travelled faster than those labeled as true. Their sample covers a minute part of the junk news universe: only those items that have been evaluated by professional fact-checkers. In contrast we propose to study the entire output of one nation's commercial junk news producers. 

The reach of political fake news appears to be over-hyped. Combining survey responses with web tracking data, Guess et al.~\cite{Guess2018} estimate that in the weeks before and after the 2016 US presidential election 1 in 4 Americans visited a fake news site, but that most fake news was consumed by a small group of conservatives. Studying the fake news audience in the US, Nelson and Taneja~\cite{nelson2018small} similarly conclude that this is a small subset of the heaviest Internet users. Political fake news is, essentially, niche content (ibid.). In contrast, the category of money-driven junk news we study aims for the largest possible audience.

The most similar to our work is the study by Fletcher et al.~\cite{Fletcher2018}. Assessing the reach of both ideologically and commercially motivated fake news in France and Italy, they downplay the problem's size, pointing out that most sites in their sample reached less than 1\% of the online population in each country.' By comparison, the most popular news websites in France (Le Figaro) and Italy (La  Repubblica) had an average monthly reach of  22.3\% and 50.9\%, respectively.'~\cite[p. 1]{Fletcher2018}. Some false news outlets, though, proved exceptionally successful:' In France, one false news outlet generated an average of over 11 million interactions per month — five times greater than more established news brands.'~\cite[p. 2]{Fletcher2018}.

This generally low level of measured engagement may be due to the study design. Fletcher et al. focussed on' outlets that consistently and deliberately publish' false news'', which we have defined elsewhere as' for-profit fabrication, politically-motivated fabrication [and] malicious hoaxes' designed to masquerade as news~\cite{Nielsen2017}.' This means that they excluded sites that publish general low-quality news, including the occasional fabricated item. Moreover, one of the Italian blacklists they used was, according to its editor, incomplete and outdated~\cite{coltelli}. Finally, employing time spent on site as a metric skews results, since users often read no more than the headline, or the abstract offered by Facebook, before hitting the share button. This latter issue was also pointed out by Coltelli~\cite{coltelli}. 

Studying data that cover one year (2017), Fletcher et al.~\cite{Fletcher2018} did not study the longitudinal development of fake news. Covering a larger time span (Jan. 2015 -- July 2018), Allcott et al.~\cite{allcott} measure the volume of Facebook user's engagements with sites known to spread false stories and compare this to developments in the reach of mainstream news sites and business and culture sites. After an initial rise in fake news engagement, this declined sharply from the beginning of 2017 onwards. During the same period, engagement numbers for the other categories they sampled remained more or less stable. The declining reach of fake news could be the result of Facebook actions against bad actors after the 2016 US elections. 

Most studies on disinformation focus on the US. The Netherlands are different in a number of respects. Actors specializing in commercially inspired political disinformation (of the kind peddled by the notorious Macedonian fake news producers targeting Trump supporters in 2016) do not exist in the Netherlands. Neither do outlets that serve nothing but fabricated news. In their capacity of third-party fact-checkers for Facebook, two of the authors reviewed hundreds of web links submitted by Dutch Facebook users as potentially ‘fake news’. Although absence of evidence does not constitute evidence of absence, we can safely assume that ‘Macedonian’ sites or sites that only publish manufactured news stories, had they existed, would have come to our attention. Moreover, neither do completely fake news sites feature in the (limited) academic literature dealing with disinformation in the Netherlands~\cite{hazenberg_micro_2018,wieringa_wie_2017} or in think tank reports~\cite{van_keulen_digitalisering_2018}, nor have they been detected by Dutch investigative journalists dealing with this topic~\cite{kist_geen_2017}.

The present study distinguishes itself from prior work in three respects: (a) it addresses commercial junk news as opposed to political junk news; (b) it addresses the phenomenon's reach on Facebook as opposed to Twitter; (c) we take a data-driven approach, including over 117 thousand published posts and the user interactions associated with these posts.

\section{Data and methods}
We compiled two seed lists of sites that we included in our sample: one of junk news sites and one of mainstream news sites. 

\subsection{Criteria for data sampling}
The criteria for including a website in the list of junk news sites were directly deduced from the definition of junk news provided above. The sites were initially brought to the attention of two of the authors in their capacity as third-party fact-checkers for Facebook. In 2017, we fact-checked more than 70 claims submitted by Facebook users; the reports were published on Nieuwscheckers.nl. Excluding the claims published on conspiracy sites (which are at least in part ideologically motivated) and on alternative health sites (adopting a different business model: some of these also make money by selling health products), left us with some 50 claims originating from commercially driven junk news sites that do not focus on one single topic. Most news items were published on multiple sites. We found that all items we checked were lacking veracity and originality: they consisted of material that was lifted from other websites and reproduced without additional research. In many cases, the sites published manufactured stories copied from foreign sources (e.g., `Oprah Winfrey (63) zwanger van eerste kind', i.e. `Oprah Winfrey (63) pregnant with first child'). In a few exceptional cases, the stories were invented by the site's editors (e.g., a story about a muslim girl from the Dutch town of Deventer who received death threats from fundamentalist muslims because she performed as a singer). 
 
By searching for other sites that had published the same news items, and by using domain information in order to identify other sites registered by the same producers, we were able to collect more junk news sites. Since many of these sites do not publish the names of their owners, we used open source information (e.g., matching Google Adsense ID numbers and public Chamber of Commerce records) to expand our list of junk news sites. We deduce the fact that the producers are not ideologically motivated from the relative absence of political content on their sites and from their personal social media use, which is also lacking political messages. We contacted seven owners and editors involved in this business, but without exception they declined to be interviewed. 

In the list of mainstream news sites we included national, well-known, general news media that have their own Facebook page. In the Dutch media landscape the set of established news media is relatively small and well-defined, consisting of national newspapers, news magazines, and news broadcasts. We only included websites that predominantly publish original, well-edited content. 

\subsection{Data download and processing}
For each domain in our junk news seed list we identified their corresponding Facebook page by crawling their homepage using Selenium~\cite{Seleniumqh.org2018} and Python, extracting the link to Facebook. For the mainstream news sites we manually identified the corresponding Facebook page. The resulting lists consist of 20 mainstream news pages and 63 junk pages, both shown in the Appendix. 

We used the Facebook API~\cite{Facebook2018} (version 2.8 accessed in the fall of 2017 for obtaining the junk news data, and version 3.0 accessed in the spring of 2018 for obtaining the mainstream news data) to download all posts published by these Facebook pages up until December 2017. The API did not return any junk news data before January 2013, most likely because the pages contained in the junk seed list were not yet active at that time. We sampled the same period for mainstream news to make both sets comparable. Thus, our sample contains all posts published by the 63 junk news pages between January 2013 and December 2017, and all posts published by the 20 mainstream news pages in that same period.

In December 2017, the Facebook API allowed us to get the unique identifiers of the users who posted a reaction or comment. Using these unique identifiers we were able to distill the number of people who interacted with a junk news post at least once. From February 6, 2017 it was no longer possible to retrieve information about user ids~\cite{QuintlySupport2018}. As a result, this information is missing for the mainstream news data.

\begin{table}[t]
\caption{General statistics about the Facebook pages included in our sample}
\begin{tabular}{lccc}
\hline
 & \# of pages & \# of posts & \# of unique users interacting with posts\\
 \hline
Mainstream news & 20 & 58,186 & Not available\\
Junk news & 63 & 58,986 & 5,285,674\\
\hline
\end{tabular}
\label{genstats}
\end{table}

\begin{figure}[t]
\includegraphics[width=\textwidth,height=.70\textheight,keepaspectratio,trim={0 3cm 0 3.5cm},clip]{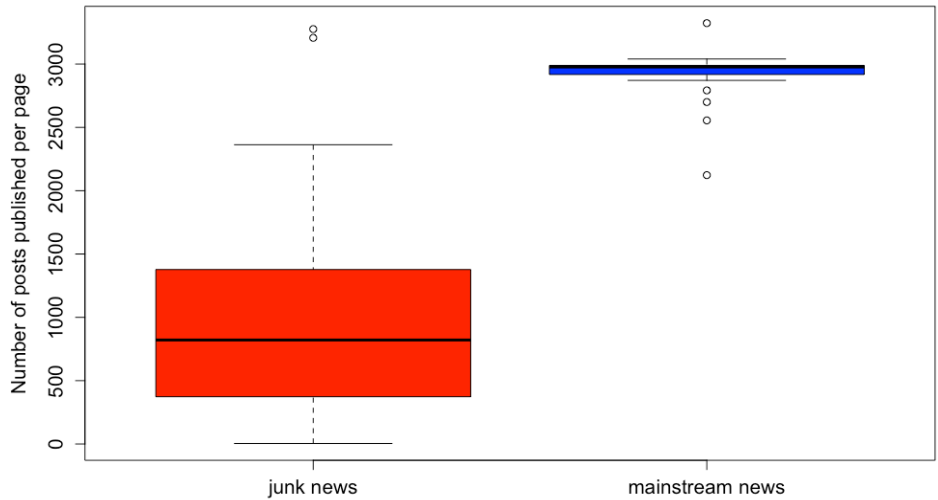}
\caption{Numbers of posts per page. The number of posts published per Facebook page, over the five-year time period, for the mainstream news pages and for the junk news pages.}
\label{postsperpage}
\end{figure}

Table~\ref{genstats} summarizes the total size of the collected data sample. Fig~\ref{postsperpage} shows the number of posts published by the individual pages, for junk news and mainstream news. The table shows that the 20 mainstream news pages have altogether published almost the same number of posts as the 63 junk news pages in the same time period. This is further illustrated by Fig~\ref{postsperpage} : each of the mainstream news pages has published more than 2000 posts in the five-year time period. Seven junk news pages have published more than 2000 posts as well, but the large majority of the junk news pages were much less active than the mainstream news pages. Table~\ref{genstats} also shows that 5,285,674 individual Facebook users interacted with a junk news post at least once. 4,055,011 individual users have added a reaction to at least one junk news post and 3,018,268 added a comment to at least one junk news post.

For each post published between January 2013 and December 2017 we retrieved the following information using the Facebook API:
\begin{itemize}
\item the publication date
\item the number of reactions to the post
\item the number of comments to the post
\item the number of times the post was shared
\end{itemize}

A `reaction' is what is commonly referred to as a `like', which can have the form of a thumbs-up, a heart, a crying emoticon, a shocked emoticon, or an angry emoticon. Reactions, comments and shares are three types of user interactions with posts on Facebook. Together they constitute the user engagement. We were unable to assess reach in terms of page views and clicks, as these are not publicly available. The publication date is needed for the longitudinal analysis of the publication activity and engagement with Facebook pages. 

We used R for the quantitative analysis of the collected data. We generated two types of statistics: statistics of the publication activity of the Facebook pages in our sample (number of posts published per month), and statistics of the user engagement with the published posts: numbers of reactions, comments and shares. 

\section{Analysis and results}
\subsection{The reach of junk news and mainstream news on Facebook}
In this section we address our first objective: to assess the total reach of junk news on Facebook, compared to mainstream news, in terms of user engagement.

\begin{table}[t]
\caption{The median, mean and standard deviation of the number of interactions (reactions, comments, and shares) per post, for mainstream news and junk news}
\begin{tabular}{lcccc}
\hline
\# of reactions & median & mean & stdev\\
\hline
Mainstream news & 42 & 362.9 & 2,082.6\\
Junk news & 70 & 429.2 & 1,739.3\\
\hline
\# of comments & median & mean & stdev\\
\hline
Mainstream news & 8 & 80.8 & 485.9\\
Junk news & 11 & 164.9 & 840.1\\
\hline
\# of shares & median & mean & stdev\\
\hline
Mainstream news & 5 & 72.5 & 1,204.5\\
Junk news & 13 & 158.3 & 1,839.3\\
\hline
\end{tabular}
\label{interactionstats}
\end{table}

\begin{figure}[th!]
\includegraphics[width=\textwidth,trim={0 3cm 0 3cm},clip]{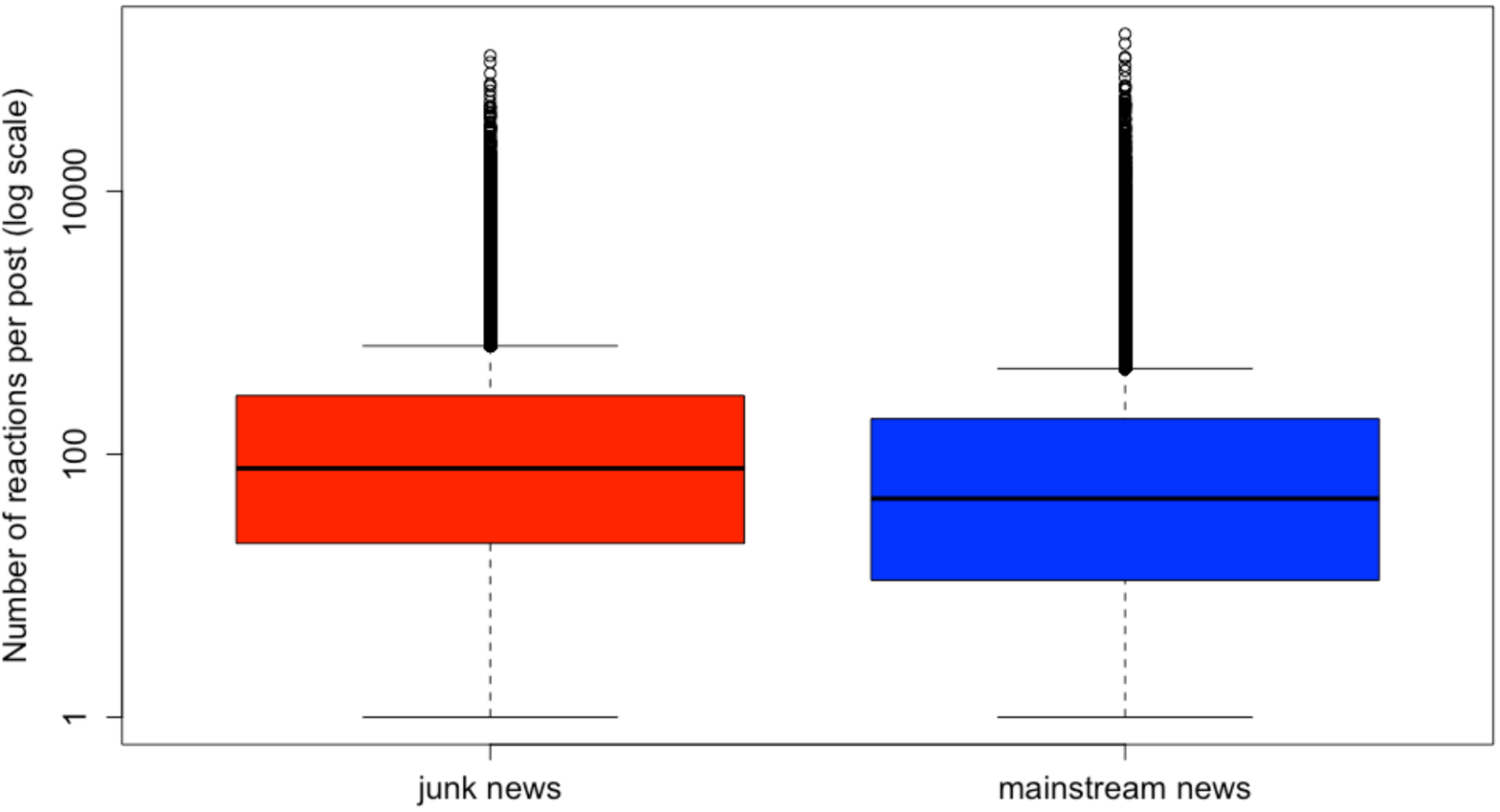}
\includegraphics[width=\linewidth,trim={0 3cm 0 3cm},clip]{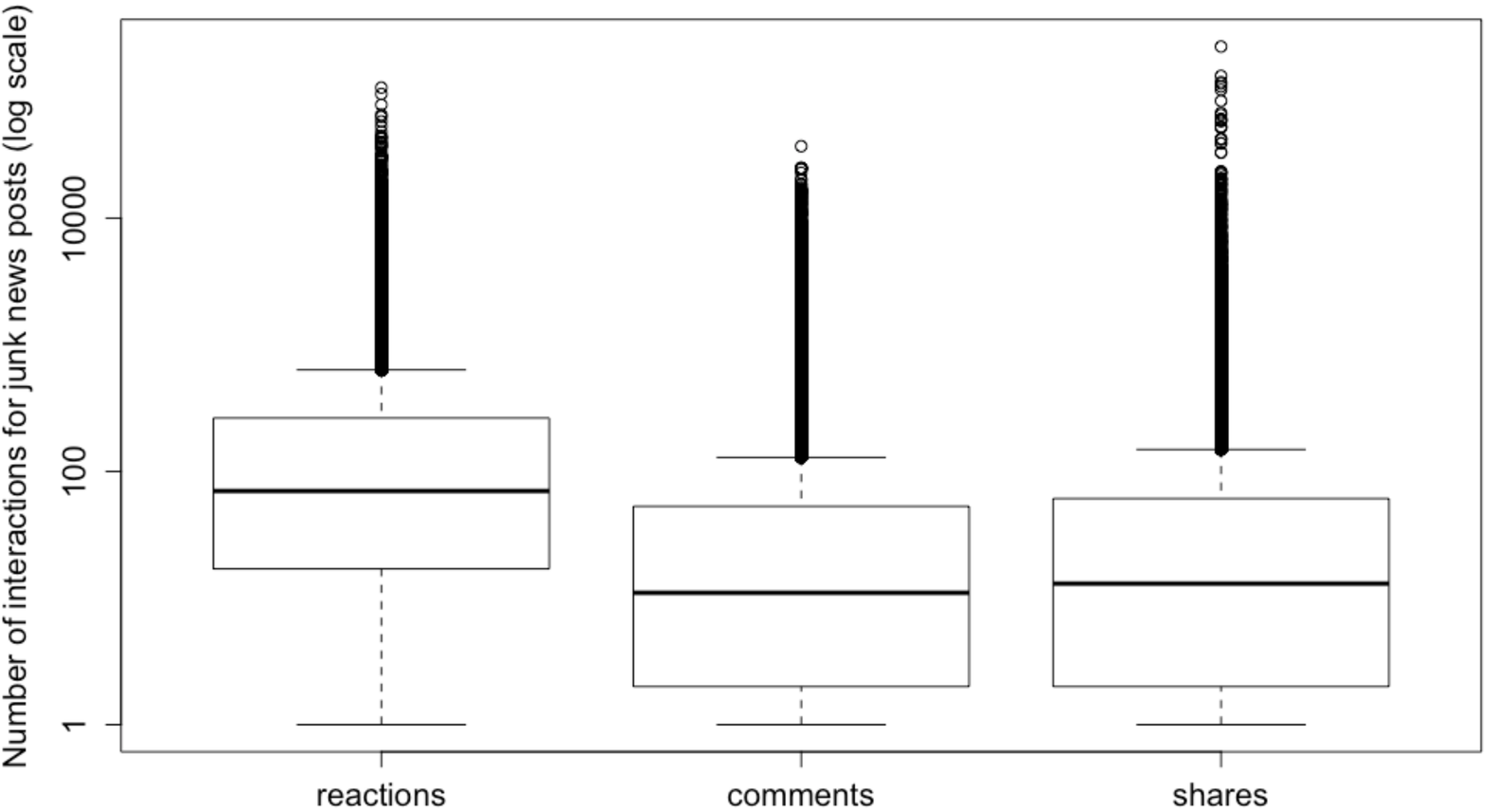}
\caption{Interactions with Facebook posts. (a) number of reactions for each news post, on a logarithmic scale; (b) the post activity for the Facebook pages, normalized by the number of pages (average number of posts published per page per month). }
\label{interactions}
\end{figure}

Table~\ref{interactionstats} lists the numbers of interactions per post over the complete five-year period, for junk news and mainstream news. An independent-samples t-test was conducted to compare the number of reactions, comments and shares on junk news and mainstream news. There was a significant difference between junk news and mainstream news for reactions, comments, and shares ($P<0.0001$ for all three comparisons). Thus, junk news on Facebook has significantly larger user engagement than mainstream news, for all engagement metrics: number of reactions, number of comments and number of shares. For example, the table shows that the median number of reactions to a junk news post is 70, compared to 42 for mainstream news. The large standard deviations and large differences between the medians and means indicate the presence of outliers. This is further illustrated by Fig~\ref{interactions}. Fig~\ref{interactions}a shows the dispersions of number of reactions for each news post, on a logarithmic scale. Fig~\ref{interactions}b shows the dispersions of numbers of reactions, comments and shares per posts, also on a logarithmic scale. The figures show large variations, with a small number of posts receiving a high number of interactions. The highest number of reactions for a single post is 107,414.

\subsection{The development of junk news and mainstream news over time}
In this section, we address our second objective: to investigate how junk news develops over time, in terms of publication activity of the junk news producers’ Facebook pages, and of user engagement with the published posts.

\subsubsection{Publication activity over time}
\begin{figure}[th!]
\includegraphics[width=1.1\textwidth,trim={2cm 2cm 0 2cm},clip]{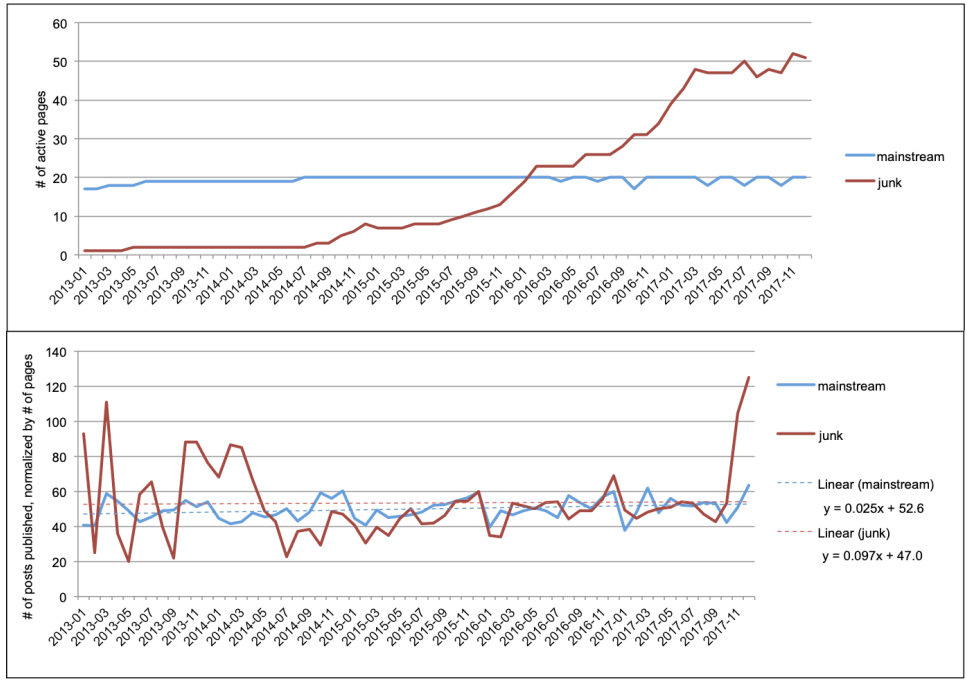}
\caption{Activity of Facebook pages. (a) the number of publishing Facebook pages per month; (b) the total post activity for the Facebook pages altogether per month. In both graphs, the red dots/line represents the junk news counts and the blue dots/line the mainstream news counts. }
\label{activity}
\end{figure}

Fig~\ref{activity} shows the publication activity over time. The average number of published posts per page per month is 50 for mainstream news (stdev=6) and 53 for junk news (stdev=21). A Mann-Whitney-Wilcoxon test indicates that the distributions in the two groups do not differ significantly ($n1 = n2 = 60, p=0.76$); thus the average publication activity per page per month is comparable between junk news pages and mainstream news pages. However, the post activity for junk news on Facebook is more irregular with a much larger standard deviation, than the post activity for mainstream news. In addition, Fig~\ref{activity}b shows a strong increase in the number of posts published per junk news page in the last three months of the measured time period (October -- December 2017). 

\subsubsection{User engagement over time}
\begin{figure}[th!]
\includegraphics[width=1.1\textwidth,trim={2cm 6.5cm 0 6.5cm},clip]{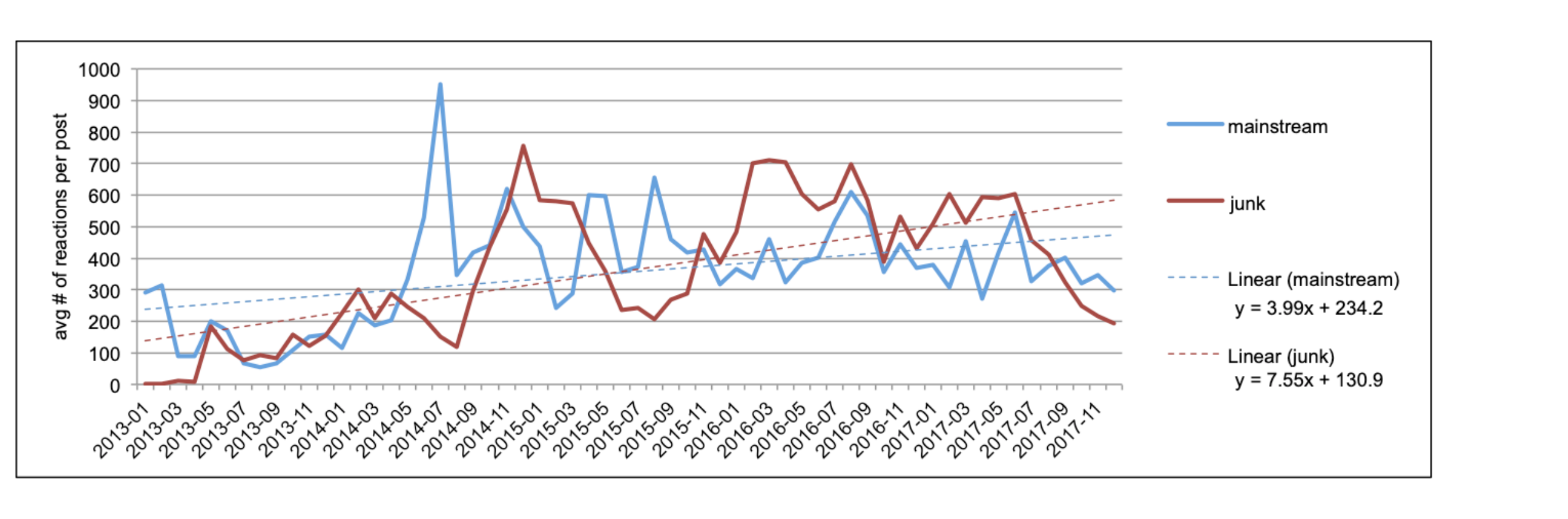}
\includegraphics[width=1.1\textwidth,trim={2cm 6.5cm 0 6.5cm},clip]{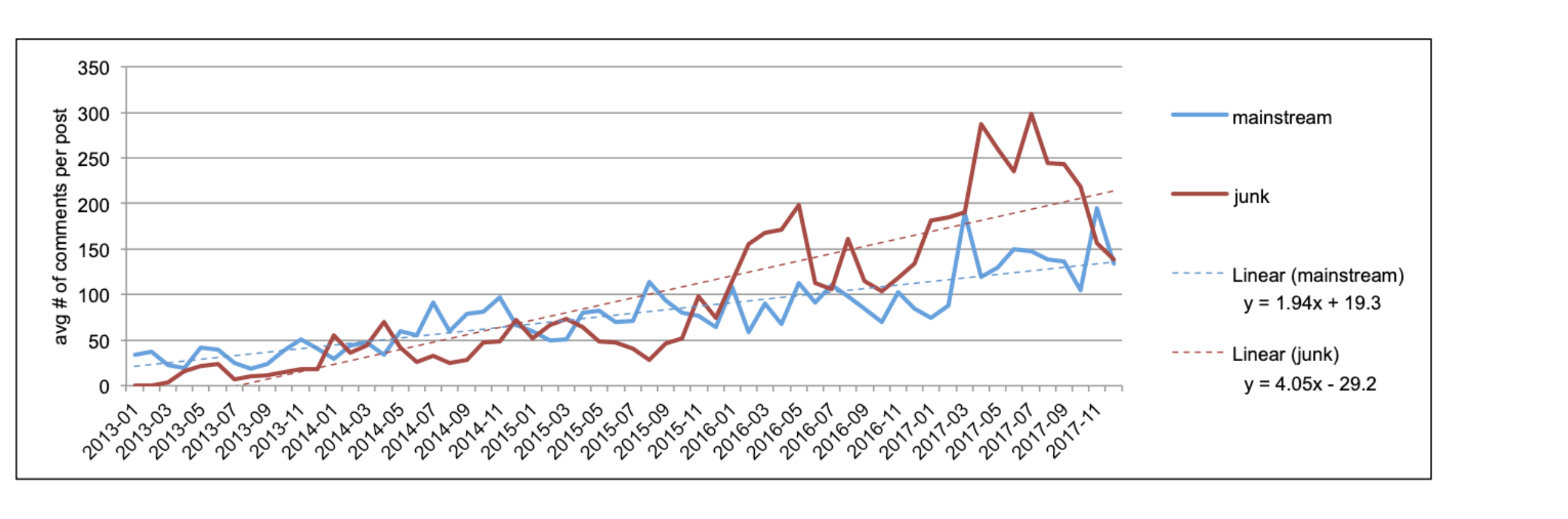}
\includegraphics[width=1.1\textwidth,trim={2cm 6.5cm 0 6.5cm},clip]{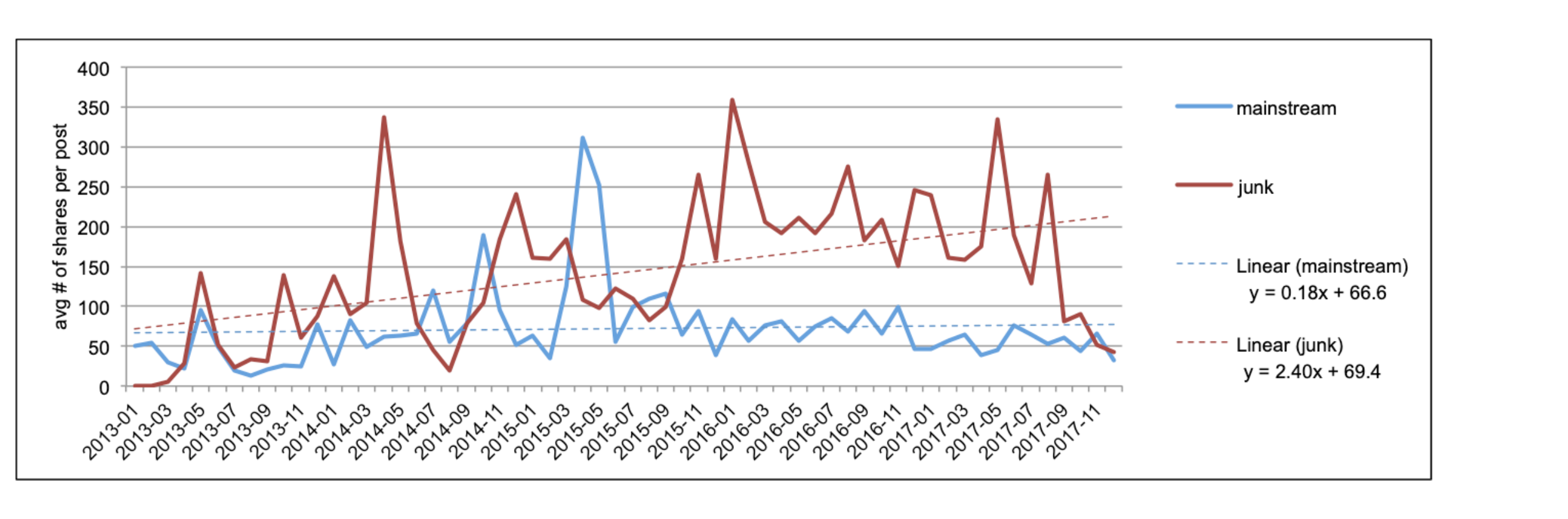}
\caption{Numbers of interactions with Facebook pages over time. (a) average number of reactions per post per month; (b) average number of comments per post per month; (c) average number of shares per post per month. In all three graphs, the red line represents the junk news counts and the blue line the mainstream news counts. Note that the y-axes of the figures have different scales.}
\label{interactionsovertime}
\end{figure}

Fig~\ref{interactionsovertime} shows the average user engagement counts per post of junk news and mainstream news over the complete time period.

Looking at the number of user interactions over time, we see that the lines for junk news and mainstream news have different peaks. We quantitatively analyzed the development of the user engagement by computing a linear least squares regression line (line of best fit) for each graph. We found that the user engagement with both types of news is growing over time, but the engagement with junk news grows faster: the slope of the trend line for reactions on junk news posts is 7.55 compared to 3.99 for mainstream news. For comments the slopes are 4.05 for junk news and 1.94 for mainstream news. For shares, the slopes are 2.40 and 0.18 respectively. An independent-samples t-test was conducted to compare the slopes of the regression lines for the change in numbers of interactions over time. There was a significant difference between junk news and mainstream news for reactions, comments, and shares. The difference between the increase of comments on junk news ($b=4.05, s.e.=39.6$) and the increase of comments on mainstream news ($b=1.94, s.e.=21.5$) was highly significant with $t(116)=5.5, p < 0.0001$. The difference between the increase of shares of junk news ($b=2.40, s.e.=77.9$) and the increase of shares of mainstream news ($b=0.18, s.e=50.5$) was highly significant with $t(116)=3.2, p = 0.0017$.

Thus, the posts published by junk news pages increasingly receive more user interactions than mainstream news. However, there is one caveat to this analysis, and that is the observation that the numbers of reactions, comments and shares for junk news pages have only decreased since the summer of 2017. This is striking because Fig~\ref{activity} has indicated that the junk news pages have become increasingly active in publishing posts in the same period, with a steep growth since September 2017.

\section{Discussion}
Quantitative research has mostly overlooked the phenomenon of money-driven junk news, focussing on junk news and fake news characterized by political content and ideological motivation. Whereas the audience for political fake news is relatively small, consisting of politically polarized, heavy media users~\cite{Guess2018,nelson2018small}, commercial junk news appears to reach the broad audience it aims for. We have shown that commercial junk news receives significantly more user interactions (reactions, comments and shares) than mainstream news on Facebook. Hiding in plain sight, this category does not strive for brand recognition or loyalty. We have demonstrated that the reach of this kind of news warrants academic attention. 

In fact, the figures we present likely underestimate the reach of junk news distributed by Facebook in the Netherlands, because we estimated reach by the number of interactions with a post. The reach of content however can be larger than the number of interactions: although it seems safe to assume that users who shared or liked a story at least read the headline, the number of Facebook users who consumed at least part of the story is probably higher than the number of people who interacted with the post~\cite[p. 65]{Tucker2018}. The data for shares, reactions, and comments are the most robust indication for the reach of junk news among Dutch Facebook users, but the number of users who must have at least scanned the headline is most likely even larger. Similarly, the number of individual users reached by the pages we sampled must be higher than the 5 Million users who added a reaction or comment to at least one of the posts. Moreover, our results show an increase in publication activity that has likely continued beyond the period we studied.

During the period covered by our data (January 2013 until December 2017) Facebook's popularity in the Netherlands has slightly grown from 9.6 million users in 2014 to 10.4 million users in 2017~\cite{Statista2018}. A similar development could be expected for its popularity as a medium for spreading content and in user engagement with the news pages. However, the user engagement with those pages show a sharper increase than the overall Facebook popularity in the same time period. Moreover, the increase of interactions with junk news is event significantly stronger than the increase of interactions with mainstream news. 

\subsection{Comparison to related studies}
Two recent studies that attempt to compare the reach of mainstream news versus fake or junk news on Facebook present findings that are less dramatic than ours. Assessing the reach of fake news in France and Italy (including money-driven fake news), Fletcher et al.~\cite{Fletcher2018} state that most sites in their sample reached less than 1\% of the online population. A French outlier however generated an average of over 11 million interactions per month, outperforming more established news brands.'

Our results deviate from Fletcher et al.'s finding that `[...] in most cases, in both France and Italy, false news outlets do not generate as many interactions as established news brands.'~\cite[p. 2]{Fletcher2018}. Our findings and our interpretation are less optimistic than those by Fletcher et al. on `fake' news in Italy and France: their findings are restricted to sites that predominantly publish completely fabricated items and their use of time spent on site as a metric neglects the fact that many users will not read beyond the headline.

Our findings also differ from those of Allcott et al.~\cite{allcott}, who compared the Facebook reach of sites known for spreading false stories with that of other news, business or culture sites. The decrease of false stories they note since early 2017 is only partly reflected in the Dutch junk news data: Our data show that junk news pages have become increasingly active in publishing posts in the second half of 2017, with a steep growth since September 2017. However, we have also observed that the numbers of reactions, comments and shares for junk news pages have decreased since the summer of 2017. We speculate that there might be a relation with Facebook's efforts to reduce the visibility of junk news on the platform  (listed in Allcott et al.~\cite{allcott}, Appendix, Table 1). In May 2017, Facebook announced that ``misinformation, sensationalism, clickbait and posts that fall outside of [their] Community Standards'' will be demoted~\cite{Facebook2018}.

Facebook data provided by the platform itself could possibly clarify this matter, but the lack of transparency about its algorithms and about the effectiveness of its actions against bad actors are a recurring obstacle for researchers in this field. As government pressure on the platforms mounts (e.g.~\cite{drozdiak}), this may change in the future. In fact, in April 2018 Facebook  and Social Science One announced a partnership in which the tech company shares data with social scientists studying fact-checking and misinformation on the platform~\cite{king,funke}. 

Dutch junk news Facebook pages frequently promote fake, i.e. fabricated, stories~\cite{BurgerP.;Pleijter}. Although these stories are not representative for the output as a whole, they can reach a sizeable audience. This is worrying, since some of these stories contain misleading health advice or false information about social groups. A completely bogus story about animal abuse by asylum seekers has been published on 13 different websites~\cite{burger}. Using Netvizz~\cite{Rieder2013}, we found that between its first publication on 17 March 2017 and 13 May 2017, the story was shared 55,292 times.

However, focusing on fabricated stories with possible social and political consequences obscures the bigger point about junk news: thriving on the core components of social media use, this highly spreadable, low-quality category of news threatens to drown out better-quality news [8].

\section{Conclusions}
We studied the reach of commercial junk news on Facebook, by analyzing 117 thousand posts published by 63 junk news pages and 20 mainstream news pages in the Netherlands. 	

In our five-year sample there is significantly larger user engagement with junk news items than with mainstream news items, for each of the three interaction metrics (reactions, comments, and shares). In terms of different people reached junk news is widespread on Facebook: 5.3 Million individual Facebook users commented or reacted on a junk news post at least once. On a total number of 10 Million Facebook users in the Netherlands this is an impressive volume of engagement.

Junk news pages have been increasingly successful in attracting user engagement over the five-year time period 2013-2017, and the increase is significantly stronger than for mainstream news. From the beginning of 2016 junk news has consistently attracted more user interactions per post than mainstream news. 

In conclusion, junk news pages are more successful than mainstream news in generating user engagement with posts. This user engagement feeds the business success for commercial junk news outlets on social media.

\bibliographystyle{splncs04}

\begin{thebibliography}{10}
\providecommand{\url}[1]{\texttt{#1}}
\providecommand{\urlprefix}{URL }
\providecommand{\doi}[1]{https://doi.org/#1}

\bibitem{allcott}
Allcott, H., Gentzkow, M., {{\&} Yu}, C.: {Trends in the Diffusion of
  Misinformation on Social Media}. arXiv preprint  (2018).
  \doi{arXiv:1809.05901}, \url{https://arxiv.org/abs/1809.05901}

\bibitem{Allcott2017}
Allcott, H., Gentzkow, M.: {Social Media and Fake News in the 2016 Election}.
  Journal of Economic Perspectives  (2017). \doi{10.1257/jep.31.2.211}

\bibitem{burger}
Burger, P.: {HOAX: Asielzoekers mishandelen puppy} (2017),
  \url{http://nieuwscheckers.nl/nieuwscheckers/hoax-asielzoekers-mishandelen-puppy/}

\bibitem{BurgerP.;Pleijter}
{Burger, P.; Pleijter}, A.: {Nieuwscheckers},
  \url{http://www.nieuwscheckers.nl}

\bibitem{coltelli}
Coltelli, M.: {The Reuters Institute for the Study of Journalism vs fake news.
  Bufale un tanto al chilo.} (2018),
  \url{http://www.butac.it/the-reuters-institute-for-the-study-of-journalism-vs-the-fake-news/}

\bibitem{DeVito2017}
DeVito, M.A.: {From Editors to Algorithms: A values-based approach to
  understanding story selection in the Facebook news feed}. Digital Journalism
  (2017). \doi{10.1080/21670811.2016.1178592}

\bibitem{drozdiak}
Drozdiak, N.: {Google, Facebook and Twitter Agree to Fight Fake News in the EU}
  (2018)

\bibitem{Facebook2018}
Facebook: {Facebook API} (2018),
  \url{https://developers.facebook.com/docs/graph-api/}

\bibitem{Fletcher2018}
Fletcher, R., Cornia, A., Graves, L., Nielsen, R.K.: {Measuring the reach of
  fake news and online disinformation in Europe} (2018),
  \url{https://reutersinstitute.politics.ox.ac.uk/sites/default/files/2018-01/Measuring
  the reach of fake news and online disinformation in Europe
  FINAL.pdf{\%}0Ahttps://reutersinstitute.politics.ox.ac.uk/our-research/measuring-reach-fake-news-and-online-disinforma}

\bibitem{funke}
Funke, D., Mantzarlis, A.: {We asked 19 fact-checkers what they think of their
  partnership with Facebook. Here's what they told us.} (2018),
  \url{https://www.poynter.org/fact-checking/2018/we-asked-19-fact-checkers-what-they-think-of-their-partnership-with-facebook-heres-what-they-told-us/}

\bibitem{Guess2018}
Guess, A., Nyhan, B., Reifler, J.: {Selective exposure to misinformation:
  evidence from the consumption of fake news during the 2016 U. S. presidential
  campaign}. (No prelo)  (2018). \doi{10.1016/j.aap.2010.08.010}

\bibitem{hazenberg_micro_2018}
Hazenberg, H.G.G., Van~den Hoven, M.J., Cunningham, S., Alfano, M.R., Asghari,
  H., Sullivan-Mumm, E.E., Ebrahimi~Fard, A., Turcios~Rodriguez, E.:
  Micro-{Targeting} and {ICT} media in the {Dutch} {Parliamentary} system:
  {Technological} changes in {Dutch} {Democracy}. Tech. rep., Delft University
  of Technology (2018)

\bibitem{house}
{House of Commons}: {Disinformation and 'fake news': Interim Report Fifth
  Report of Session 2017–19} (2018),
  \url{https://publications.parliament.uk/pa/cm201719/cmselect/cmcumeds/363/363.pdf}

\bibitem{howard}
Howard, P.N., Bolsover, G., Kollanyi, B., Bradshaw, S., Neudert, L.M.: {Junk
  News and Bots during the U.S. Election: What Were Michigan Voters Sharing
  Over Twitter? Data Memo 2017.1} (2017)

\bibitem{Howard2017}
Howard, P.N., Bradshaw, S., Kollanyi, B., Desigaud, C., Bolsover, G.: {Junk
  News and Bots during the French Presidential Election: What Are French Voters
  Sharing Over Twitter?} In: COMPROP DATA MEMO (2017)

\bibitem{Howard2018}
Howard, P.N., Woolley, S., Calo, R.: {Algorithms, bots, and political
  communication in the US 2016 election: The challenge of automated political
  communication for election law and administration}. Journal of Information
  Technology and Politics  (2018). \doi{10.1080/19331681.2018.1448735}

\bibitem{humprecht2018fake}
Humprecht, E.: {Where ‘fake news' flourishes: a comparison across four
  Western democracies}. Information, Communication {\&} Society pp. 1--16
  (2018)

\bibitem{king}
King, G., Persily, N.: {A New Model for Industry-Academic Partnerships} (2018),
  \url{https://gking.harvard.edu/files/gking/files/partnerships.pdf}

\bibitem{kist_geen_2017}
Kist, R., Zanthingh, P.: Geen grote rol nepnieuws in aanloop naar verkiezingen.
  NRC  (Mar 2017),
  \url{https://www.nrc.nl/nieuws/2017/03/06/fake-news-nee-zo-erg-is-het-hier-niet-7144615-a1549050}

\bibitem{mosseri}
Mosseri, A.: {Building a Better News Feed for You. Facebook Newsroom} (2016),
  \url{https://newsroom.fb.com/news/2016/06/building-a-better-news-feed-for-you/}

\bibitem{Narayanan2018}
Narayanan, V., Barash, V., Kollanyi, B., Neudert, L.M., Howard, P.N.:
  {Polarization, Partisanship and Junk News Consumption over Social Media in
  the US}. arXiv preprint  (2018), \url{https://arxiv.org/abs/1803.01845}

\bibitem{nelson2018small}
Nelson, J.L., Taneja, H.: {The small, disloyal fake news audience: The role of
  audience availability in fake news consumption}. new media {\&} society p.
  1461444818758715 (2018)

\bibitem{newman}
Newman, N., Fletcher, R., Kalogeropoulos, A., Levy, D.A.L., {Kleis Nielsen},
  R.: {Reuters Institute Digital News Report 2018} (2018),
  \url{http://media.digitalnewsreport.org/wp-content/uploads/2018/06/digital-news-report-2018.pdf?x89475}

\bibitem{Nielsen2017}
Nielsen, R.K., Graves, L.: {"News you don't believe": Audience perspectives on
  fake news} (2017). \doi{10.2174/1874205X01105010068}

\bibitem{powers}
Powers, S., Kounalakis, M.: {Can Public Diplomacy Survive the Internet? Bots,
  Echo Chambers, and Disinformation} (2017),
  \url{https://www.state.gov/documents/organization/271028.pdf}

\bibitem{Rieder2013}
Rieder, B.: {Studying Facebook via Data Extraction: The Netvizz Application}.
  In: Proceedings of WebSci '13 - 5th Annual ACM Web Science Conference (2013).
  \doi{10.1145/2464464.2464475}

\bibitem{Roozenbeek2018}
Roozenbeek, J., van~der Linden, S.: {The fake news game: actively inoculating
  against the risk of misinformation} (2018).
  \doi{10.1080/13669877.2018.1443491}

\bibitem{Seleniumqh.org2018}
Seleniumqh.org: {SeleniumQH browser automation} (2018),
  \url{https://www.seleniumhq.org/}

\bibitem{senecat2017}
S{\'{e}}n{\'{e}}cat, A.: {Sushis, vaccins et viande humaine : le «
  palmar{\`{e}}s » des fausses infos} (2017),
  \url{https://www.lemonde.fr/le-blog-du-decodex/article/2017/09/08/sushis-vaccins-et-viande-humaine-le-palmares-des-fausses-infos{\_}5182743{\_}5095029.html}

\bibitem{silverman}
Silverman, C., Feder, J.L., Cvetkovska, S., Belford, A.: {Macedonia's Pro-Trump
  Fake News Industry Had American Links, And Is Under Investigation For
  Possible Russia Ties. Buzzfeed.} (2018),
  \url{https://www.buzzfeednews.com/article/craigsilverman/american-conservatives-fake-news-macedonia-paris-wade-libert}

\bibitem{Statista2018}
Statista: {Number of Facebook users in the Netherlands 2014-2018 | Statistic}
  (2018),
  \url{https://www.statista.com/statistics/283635/netherlands-number-of-facebook-users/}

\bibitem{QuintlySupport2018}
Support, Q.: {Changes to Facebook's API - February 6th, 2018} (2018),
  \url{https://support.quintly.com/hc/en-us/articles/115004414274-Changes-to-Facebook-s-API-February-6th-2018}

\bibitem{Tandoc2018}
Tandoc, E.C., Lim, Z.W., Ling, R.: {Defining “Fake News”: A typology of
  scholarly definitions} (2018). \doi{10.1080/21670811.2017.1360143}

\bibitem{Tucker2018}
Tucker, J.A., Guess, A., Barbera, P., Vaccari, C., Siegel, A., Sanovich, S.,
  Stukal, D., Nyhan, B.: {Social Media, Political Polarization, and Political
  Disinformation: A Review of the Scientific Literature}. SSRN  (2018).
  \doi{10.2139/ssrn.3144139}

\bibitem{van_keulen_digitalisering_2018}
Van~Keulen, I., Korthagen, I., Diederen, P., van Boheemen, P.: Digitalisering
  van het nieuws – {Online} nieuwsgedrag, desinformatie en personalisatie in
  {Nederland}. Tech. rep., Rathenau Instituut, The Hague, the Netherlands
  (2018),
  \url{https://www.rathenau.nl/sites/default/files/2018-05/Digitalisering%20van%20het%20nieuws.pdf}

\bibitem{Vargo2018}
Vargo, C.J., Guo, L., Amazeen, M.A.: {The agenda-setting power of fake news: A
  big data analysis of the online media landscape from 2014 to 2016}. New Media
  and Society  (2018). \doi{10.1177/1461444817712086}

\bibitem{venturini}
Venturini, T.: {From Fake to Junk News, the Data Politics of Online Virality}.
  In: D. Bigo, E. Isin, {\&} E. Ruppert (Eds.), Data Politics: Worlds,
  Subjects, Rights. London: Routledge (2019),
  \url{http://www.tommasoventurini.it/wp/wp-content/uploads/2018/10/Venturini_FromFakeToJunkNews.pdf}

\bibitem{Vosoughi2018}
Vosoughi, S., Roy, D., Aral, S.: {The spread of true and false news online}.
  Science  \textbf{359}(6380),  1146--1151 (2018).
  \doi{10.1126/science.aap9559},
  \url{http://www.sciencemag.org/lookup/doi/10.1126/science.aap9559}

\bibitem{wardle}
Wardle, C., Derakhshan, H.: {Information Disorder: Toward an interdisciplinary
  framework for research and policymaking} (2017)

\bibitem{wieringa_wie_2017}
Wieringa, M., De~Winkel, T., Lewis, C.: Wie is de waakhond op sociale media.
  Tech. rep., Nederlands Genootschap van Hoofdredacteuren. Nicht … (2017)

\end{thebibliography}

\section*{Appendix}

\begin{table}[h]
\caption{List of 20 Dutch mainstream news Facebook pages included in our sample}
\begin{tabular}{lll}
\hline
URL & Name & Category\\
\hline
facebook.com/190231243842 & De Groene Amsterdammer & News magazine\\
facebook.com/103652819717294 & HP/De Tijd & News magazine\\
facebook.com/ad.nl & Algemeen Dagblad & National newspaper\\
facebook.com/bnr.nieuwsradio & BNR Nieuwsradio & News broadcast\\
facebook.com/decorrespondent & De Correspondent & Online news magazine\\
facebook.com/elsevierweekblad & Elsevier  & News magazine\\
facebook.com/geenstijlnl & Geenstijl & Online news magazine\\
facebook.com/hetfd & Het Financieele Dagblad & National newspaper\\
facebook.com/metro & Metro & National newspaper\\
facebook.com/nos & NOS & News broadcast\\
facebook.com/nporadio1 & NPO Radio 1 & News broadcast\\
facebook.com/nrc & NRC & National newspaper\\
facebook.com/nu.nl & Nu.nl & Online newspaper\\
facebook.com/paroolnl & Het Parool & National newspaper\\
facebook.com/refdag & Reformatorisch Dagblad & National newspaper\\
facebook.com/rtlnieuws & RTL Nieuws & News broadcast\\
facebook.com/telegraaf & Telegraaf & National newspaper\\
facebook.com/tponl & The Post Online & Online newspaper\\
facebook.com/trouw.nl & Trouw & National newspaper\\
facebook.com/volkskrant & Volkskrant & National Newspaper\\
\hline
\end{tabular}
\end{table}

\begin{longtable}{ll}
\caption{List of 63 Dutch junk news pages included in our sample}\\
\hline
URL & Name (if not in URL)\\
\hline
facebook.com/106727739817828 & ROOS\\
facebook.com/1213051412048021 & Suri.nu\\
facebook.com/1315496065181384 & Ongelooflijk Favorieten\\
facebook.com/1364984373581269 & Originele Ideeën\\
facebook.com/1567673943543704 & Dierenvriend\\
facebook.com/1568382686520046 & Videodump\\
facebook.com/1661870274079781 & Kookfans\\
facebook.com/1721172708118846 & Viralfilmpje\\
facebook.com/1785829101687628 & LEESR\\
facebook.com/187487161609007 & Viraal.co\\
facebook.com/1931812020438753 & Dames\&Heren\\
facebook.com/195089190827932 & Tips en Weetjes\\
facebook.com/386130775068830 & Dol op wilde dieren\\
facebook.com/410280812497912 & Doedatzelf\\
facebook.com/572385656268218 & Viralfy\\
facebook.com/662965343876026 & Ghettoland\\
facebook.com/688141944701187 & \\
facebook.com/architectdistrict & \\
facebook.com/arkinoh & \\
facebook.com/bekijkdezevideo & \\
facebook.com/bengbengnl & \\
facebook.com/blijfpositiefcom & \\
facebook.com/brawnl & \\
facebook.com/breekdedag & \\
facebook.com/curioctopus.nl & \\
facebook.com/dagelijks.nu & \\
facebook.com/deelze.nl & \\
facebook.com/ditisgeniaal & \\
facebook.com/echte.mannen.wereld & \\
facebook.com/eetradar & \\
facebook.com/feitjesenweetjes.nl & \\
facebook.com/foodviral & \\
facebook.com/forestfeed & \\
facebook.com/gezondeideetjes & \\
facebook.com/grappig.co & \\
facebook.com/hetdelenwaard & \\
facebook.com/indrukwekkend.nu & \\
facebook.com/kingbreaknl & \\
facebook.com/leeftips.nl & \\
facebook.com/leeshetnu.nl & \\
facebook.com/leeshetpuntnu & \\
facebook.com/lhoriginale & \\
facebook.com/lijpeshitt & \\
facebook.com/livekijken & \\
facebook.com/luidt.nl & \\
facebook.com/newsnernederlands & \\
facebook.com/nieuwsco & \\
facebook.com/ongelofelijk.eu & \\
facebook.com/pranksternl & \\
facebook.com/secretmancave & \\
facebook.com/straatvidsnl & \\
facebook.com/toptrendingnl & \\
facebook.com/trendnieuws & \\
facebook.com/vandaagviraal & \\
facebook.com/verhalen.co & \\
facebook.com/viraaltjes & \\
facebook.com/viraalvandaag & \\
facebook.com/viral2day.nl & \\
facebook.com/viralnextnieuws & \\
facebook.com/viralsonline1 & \\
facebook.com/viraltje & \\
facebook.com/volgendevideo & \\
facebook.com/zelfmaakideetjes & \\
\hline
\end{longtable}

\end{document}